\documentclass{aa}
\usepackage{graphicx}
\newcommand{\degree}{\mbox{$^{\circ}$}}               
\newcommand{\magnit}[2]{\mbox{$\mbox{\rm #1}^{\mbox{\rm\tiny m}}%
     \!\!\!.\!\,\, \mbox{\rm #2}$}}                   
\newcommand{\magap}[1]{\mbox{$\mbox{\rm #1}^{\mbox{\rm\tiny m}}$}} 
\newcommand{\filter}[1]{\mbox{\it #1\/}}              
\newcommand{\colour}[2]{\mbox{(\it #1$ - $#2\/})}     
\newcommand{\oversim}[2]{\lower0.5ex\vbox{\baselineskip=0pt\lineskip=0.2ex
     \ialign{$\mathsurround=0pt #1\hfil##\hfil$\crcr#2\crcr\sim\crcr}}}
\newcommand{\etal}{\mbox{\hbox{et\,al.}}}         
\newcommand{\eg}{\mbox{\hbox{e.g.,}}}             
\begin{document}
\title{Search for nearby stars among proper motion stars
       selected by optical-to-infrared photometry.}

\subtitle{II\@. Two late M dwarfs within 10 pc
\thanks{Based on observations collected with the ESO 3.6-m/EFOSC2 at the 
European Southern Observatory, La Silla, Chile (ESO programme 68.C-0664).}}

\author{M. J. McCaughrean \inst{} \and
        R.-D. Scholz\inst{} \and
        N. Lodieu \inst{}}

\offprints{M. J. McCaughrean}

\institute{Astrophysikalisches Institut Potsdam, 
           An der Sternwarte 16, 14482 Potsdam, Germany\\
           \email{mjm@aip.de, rdscholz@aip.de, nlodieu@aip.de}}

\date{Received ...; accepted ...}

\titlerunning{Search for nearby stars. II\@. Two late M dwarfs within 10 pc}
\authorrunning{McCaughrean, Scholz, \& Lodieu}

\abstract{
We have identified two late M dwarfs within 10\,pc of the Sun, by
cross-correlating the Luyten NLTT catalogue of stars with proper motions 
larger than 0.18\,arcsec/yr, with objects lacking optical identification in the 
2MASS data base. The 2MASS photometry was then combined with improved optical
photometry obtained from the SuperCOSMOS Sky Surveys. The two objects 
(LP\,775-\,31 and LP\,655-\,48) have extremely red optical-to-infrared colours 
(\colour{R}{K$_s$}$\sim$\magap{7}) and very bright infrared magnitudes 
(\filter{K$_s$}$<$\magap{10}): follow-up optical spectroscopy with the 
ESO 3.6-m telescope gave spectral types of M8.0 and M7.5 dwarfs, 
respectively. Comparison of their near-infrared magnitudes with the absolute 
magnitudes of known M8 and M7.5 dwarfs with measured trigonometric 
parallaxes yields spectroscopic distance estimates of $6.4\pm1.4$\,pc and 
$8.0\pm1.6$\,pc for LP\,775-\,31 and LP\,655-\,48, respectively. In contrast,
Cruz \& Reid (2002) recently determined spectral types of M6 for both objects,
and commensurately larger distances of $11.3\pm 1.3$\,pc and $15.3\pm 2.6$\,pc.
LP\,655-\,48 is also a bright X-ray source (1RXS~J044022.8-053020). 
With only a few late M~dwarfs previously known within 10\,pc, these two 
objects represent an important addition to the census of the Solar 
neighbourhood.
}
\maketitle

\section{Introduction}
The census of the solar neighbourhood remains very incomplete. From the 
statistics of the Catalogue of Nearby Stars (Gliese \& Jahrei\ss{} 
\cite{gliese91}), one can infer that about 30\% of all stars (and probably 
an even larger fraction of the substellar brown dwarfs) within 10\,pc are 
as yet undetected. These sources are important however, as the detailed 
observation of very nearby stars and brown dwarfs is one of the main starting 
points for investigations of the star formation process, the stellar and 
substellar luminosity function, and the initial mass function. Furthermore, 
future missions for the detection of extrasolar planets (\eg{} SIM, TPF, 
Darwin) will concentrate on very nearby stars in order to reveal Earth-like 
planets. As a result, several major efforts are underway to provide a more 
detailed census of the solar neighbourhood, including the RECONS and NStars 
projects, out to 10 and 25\,pc, respectively (www.chara.gsu.edu/RECONS and
nstars.arc.nasa.gov). Recent discoveries in the immediate solar neighbourhood 
($<$\,15\,pc) include early and mid M~dwarfs 
(Jahrei{\ss} \etal{} \cite{jahreiss01};
Scholz \etal{} \cite{scholz01}; 
Phan-Bao \etal{} \cite{phanbao01};
Scholz \etal{} \cite{scholz02a};
Reid \& Cruz \cite{reid02};
Reid \etal{} \cite{reid02a};
Cruz \& Reid \cite{cruz02};
Reyl\'e \etal{} \cite{reyle02}),
late~M  and L~dwarfs 
(Gizis \etal{} \cite{gizis00};
Kirkpatrick \etal{} \cite{kirkpatrick00};
Reid \etal{} \cite{reid00};
Delfosse \etal{} \cite{delfosse01};
Reid \& Cruz \cite{reid02};
Reid \etal{} \cite{reid02a};
Cruz \& Reid \cite{cruz02};
Lodieu \etal{} \cite{lodieu02}),
and cool white dwarfs
(Reid \etal{} \cite{reid01}; 
Scholz \etal{} \cite{scholz02b}).

All of these objects show a proper motion larger than the NLTT (New Luyten Two 
Tenths catalogue; Luyten \cite{luyten7980}) limit of 0.18\,arcsec/yr, and 
indeed, many of the early and mid M~dwarfs were selected from the NLTT\@. More 
than 20 years after its publication, the NLTT is still an important source in 
the search for nearby stars and by far not yet fully exploited. North of
$\delta$=$-33\degree$, the NLTT catalogue is based on Palomar Schmidt plates 
and has a limiting magnitude of $m_{\rm pg}$$\sim$\magnit{20}{5} (see,
\eg{} Reid \& Cruz \cite{reid02}). South of $\delta$=$-33\degree$, the 
limiting magnitude of the NLTT is only $m_{\rm pg}$$\sim$\magnit{15}{5}, and 
thus new high proper motion surveys have been started using UK Schmidt 
Telescope (UKST) and ESO Schmidt plates (Scholz \etal{} \cite{scholz00}; 
Ruiz \etal{} \cite{ruiz01}).

Many recent additions to the census of late M~and early L~dwarfs within 15\,pc
were found in the near-infrared sky surveys 2MASS (Gizis \etal{} 
\cite{gizis00}; Kirkpatrick \etal{} \cite{kirkpatrick00}; Reid \etal{} 
\cite{reid00}) and DENIS (Delfosse \etal{} \cite{delfosse01}). Nevertheless, 
all these objects could in principle have been selected directly from the 
NLTT and/or modern Schmidt plate measurements, and in fact, all of them were 
subsequently identified as high proper motion objects in the NLTT or from 
Schmidt plates.

With the recent discovery of three nearby L~dwarfs (one with a spectroscopic 
distance estimate of less than 15\,pc) in the southern sky (Lodieu \etal{}
\cite{lodieu02}), we have demonstrated that high proper motion surveys can 
successfully contribute to the completion of the census of the solar 
neighbourhood at the stellar/substellar boundary. Furthermore, the combination 
of old and new proper motion catalogues with modern near-infrared sky surveys
is a very effective tool in the search for nearby red dwarfs. The NLTT and 
2MASS have been used to discover an M6.5 dwarf at $\sim$\,6\,pc (Scholz
\etal{} \cite{scholz01}), and five new sources were identified within 15\,pc
(and one closer than 10\,pc) from a new proper motion survey combined with 
DENIS data (Reyl\'e \etal{} \cite{reyle02}). Most recently, Reid \& Cruz 
(\cite{reid02}), Reid \etal{} (\cite{reid02a}), and Cruz \& Reid 
(\cite{cruz02}) have reported the discovery of many additions to the immediate 
solar neighbourhood from NLTT stars identified in the 2MASS, including five 
objects within 10\,pc, 

In this paper, we describe the discovery of two very nearby ($<10$\,pc) late 
M~dwarfs in the southern sky, also found in a combined search of the NLTT and 
2MASS\@. The same two objects were also recently identified by Cruz \&
Reid (2002) with different results, as described below.

\section{Re-identification and photometry of NLTT stars}
In order to identify the high proper motion stars of the NLTT catalogue 
(Luyten \cite{luyten7980}) in the 2MASS data base, a re-identification of the 
objects in the optical is first necessary. As recently shown by Gould \& Salim 
(\cite{gould02}), Luyten measured positions for the majority of the stars in 
his catalogue to 1 arcsec accuracy, but recorded the results to an accuracy of 
only 1 second of time in right ascension and 6 arcsec in declination. In 
addition, there is a long tail of positional errors ranging from tens to 
hundreds of arcsec. Thus, re-identification requires a visual inspection of 
multi-epoch data obtained from scans of photographic Schmidt plates. 

Also, as the stars are re-identified, improved photographic photometry can be 
obtained to compensate for the relatively low photometric accuracy of the NLTT 
catalogue (Scholz \etal{} \cite{scholz00}; Gould \& Salim \cite{gould02}; Reid 
\& Cruz \cite{reid02}). In the northern sky, the USNO A2.0 catalogue (Monet 
\etal{} \cite{monet98}) can be used, but south of $\delta$=$-20$\degree{}, most 
NLTT stars are not included, due to the epoch difference between the Schmidt 
plates and the object matching radius used in the construction of the catalogue.

Thus here we have used the SuperCOSMOS Sky Survey (hereafter SSS: 
Hambly \etal{} 
\cite{hambly01a,hambly01b,hambly01c}), a recent scan of the entire southern sky 
made from UKST and ESO Schmidt plates in three passbands (\filter{B$_J$}, 
\filter{R}, and \filter{I}). The SSS data are available in the form of images
which can be used for the re-identification of NLTT stars, and object 
catalogues, including accurate photographic photometry from each of the plates. 
The epoch differences between the plates vary from field to field, with 
typically a maximum of 15--20 years between the UKST \filter{B$_J$} and 
\filter{R} plates. The quality of and easy access to the SSS data make it
well suited to the re-identification and photometry of NLTT stars, and we
have also used these data to conduct a new high proper motion survey which
has led to the discovery of a new wide pair of cool white dwarfs in the
solar neighbourhood (Scholz \etal{} \cite{scholz02b}) and three nearby 
L~dwarfs (Lodieu \etal{} \cite{lodieu02}). 

\begin{table*}
\caption[]{Astrometry and photometry from the SSS and 2MASS\@. The proper 
motion for each object is based on five positions: the \filter{B$_J$}, 
\filter{R}, and \filter{I} plates measured within the SSS, the 2MASS position, 
and the USNO A2.0 early epoch position. The coordinates are given for the most 
recent epoch, which is 2MASS for LP\,775-\,31 and the SSS \filter{I} plate 
for LP\,655-\,48.}
\label{tab:sss2mass}
\begin{center}
\begin{tabular}{ccccrcccccc}
\hline
Name & $\alpha, \delta$ & Epoch & $\mu_{\alpha}\cos{\delta}$ & $\mu_{\delta}$ & 
\filter{B$_J$} & \filter{R} & \filter{I} & 
\filter{J}     & \filter{H} & \filter{K$_s$} \\
LP   & (J2000) & & \multicolumn{2}{c}{mas/yr} & 
\multicolumn{3}{c}{(SSS)} & \multicolumn{3}{c}{(2MASS)} \\ \hline
775-\,31 & 04 35 16.14~~$-$16 06 57.5 & 1998.90 & $+160 \pm 4$ & $+305 \pm 4$ & 
18.85 & 16.35 & 12.36 & 10.40 & 9.78 & 9.34 \\
655-\,48 & 04 40 23.33~~$-$05 30 07.9 & 2001.80 & $+335 \pm 2$ & $+131 \pm 2$ & 
18.85 & 16.50 & 13.17 & 10.68 & 9.99 & 9.56 \\
\hline
\end{tabular}
\end{center}
\end{table*} 

\begin{figure*}
\centering
\includegraphics[angle=-90,width=16.7cm]{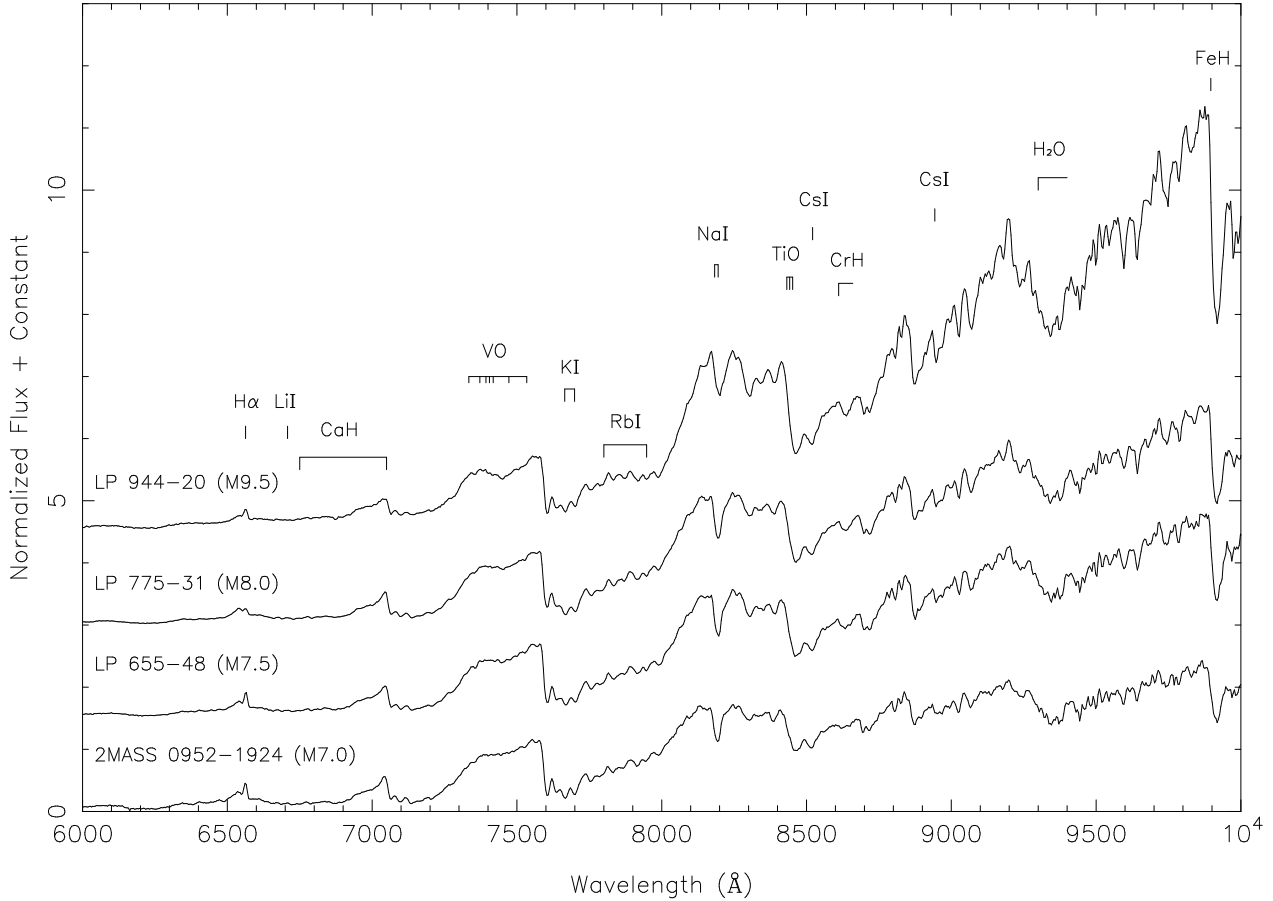}
\caption{ESO 3.6-m/EFOSC2 spectra of 
LP\,775-\,31 and LP\,655-\,48, compared with those of LP\,944-\,20, a 
known M9.5 brown dwarf (Tinney \cite{tinney96,tinney98}; Reid \etal{} 
\cite{reid02b}), and the M7.0 dwarf 2MASS~0952219$-$192431 (Gizis \etal{}
\cite{gizis00}). An arbitrary constant has been used to separate the
spectra. The location of features typical of late-type stars are labelled, 
including metal oxide and hydride absorption bands, atomic absorption lines, 
lithium absorption, and H$\alpha$ emission.
}
\label{fig:4spec}
\end{figure*}

\section{Selecting objects from the 2MASS}
To yield a list of candidate nearby red dwarfs in the southern sky, we first
searched the 2MASS for bright (\filter{K$_s$}$<$\magap{11}) southern stars 
without optical counterparts. As the USNO A2.0 catalogue is used to provide 
optical data for 2MASS stars, the large epoch difference between the A2.0 and 
2MASS observations ensures that high proper motion stars usually end up without 
an optical counterpart in the 2MASS catalogue. 

The resulting $>$10$^6$ 2MASS sources without optical counterparts were then
cross-correlated with the NLTT catalogue, using a search radius of 60 arcsec.
In practice, most matches were found in the region 
$-33$\degree$<$$\delta$$<$0\degree{} due to the limited NLTT sensitivity south 
of that: of this region, roughly 90\% is covered in the publicly-available
2MASS database. For any objects thus identified, a very large 
optical-to-infrared colour index of \colour{m$_r$}{K$_s$}$>$\magap{6} was used 
as a further selection criterion, where \filter{m$_r$} is the rough red 
magnitude estimate in the NLTT catalogue. 

Only $\sim$\,170 matches remained after this selection, and these were then
re-identified and checked using SSS images, and \filter{B$_J$}, \filter{R}, and
\filter{I} photometry was obtained from the SSS catalogues. The final list of 
candidates consisted of just 9 NLTT entries with 
\colour{R}{K$_s$}$>$\magnit{6}{5}. Six of these were known M7--M9.5
dwarfs and one object was not visible during our spectroscopic follow-up 
observations. The two remaining candidates, LP\,775-\,31 and LP\,655-\,48,
were among the brightest 2MASS sources in the sample of 9 objects. Other 
sources with comparable brightness (\filter{K$_s$}$\sim$\magnit{9}{5}) include 
the known brown dwarf LP\,944-\,20 (Tinney \cite{tinney96,tinney98}) and the 
recently discovered M9 dwarf DENIS-P\,J104814.7$-$395606.1 (Delfosse \etal{}
\cite{delfosse01}).

The full astrometric and photometric data for LP\,775-\,31 and LP\,655-\,48
are given in Table~\ref{tab:sss2mass}. It is worth noting that LP\,655-\,48 
is also identified with a bright X-ray source (1RXS~J044022.8-053020;
Voges \etal{} \cite{voges99}).

\section{Optical classification spectroscopy}
Optical classification spectroscopy was obtained for the two candidate nearby
sources using EFOSC2 mounted on the ESO 3.6-m telescope on La Silla on 22 
November 2001. The conditions were photometric and the seeing $\sim$\,0.6--0.8 
arcsec FWHM\@. The spectrograph uses a 2048$\times$2048 pixel Loral/Lesser 
CCD with a pixel size of 0.157\,arcsec, and has a useable field-of-view of 
5.2$\times$5.2 arcmin. A 1 arcsec slit was used with Grism 12 covering 
6000--10300\AA{} at R\,$\sim$\,600. Single exposures of 240\,s were
obtained for each of the two candidates, and of 200\,s for each of two 
comparison late M dwarfs (LP\,944-\,20 and 2MASS\,0952219$-$192431). 

Data reduction was standard, including subtraction of an averaged dark frame 
and division by an internal quartz flat field to remove fringing above 
8000\AA{}. Wavelength calibration was made using He and Ar lines over the 
whole wavelength range, and flux calibration was achieved using an averaged 
sensitivity function determined from several exposures of the 
spectrophotometric standard stars EG21 and LTT1788 taken during the night.

After extraction of the spectra, the PC3 index defined by Mart\'{\i}n \etal{} 
(\cite{martin99}) was used to determine spectral types of M6.5, M7.5, M8.0, 
and M9.5 for 2MASS\,0952219$-$192431, LP\,655-\,48, LP\,775-\,31, and 
LP\,944-\,20, respectively (see Fig.~\ref{fig:4spec}). Spectral types 
determined using the classification scheme of Kirkpatrick \etal{} 
(\cite{kirkpatrick99}) were consistent to within the half subclass errors.
The spectral types found for the comparison objects are in good agreement 
with previously published values: the M9.5 determined for LP\,944-\,20 is 
consistent with the recent value given by Reid \etal{} (\cite{reid02b}),
while Tinney (\cite{tinney96}) gave it as M9 on its discovery.

Thus for the two new sources, LP\,655-\,48 and LP\,775-\,31, we adopt spectral
types M7.5 and M8.0, respectively. We note that Cruz \& Reid (2002) gave
M6 for both stars, but the direct spectroscopic evidence presented here
strongly favours the later types we have adopted.

\section{Determination of spectroscopic distances}
The data of Reid \etal{} (\cite{reid02b}) yield trigonometric distances and
absolute \filter{M$_J$}, \filter{M$_H$}, \filter{M$_{K_s}$} magnitudes for three
M8 and four M7.5 dwarfs with 2MASS measurements. The abbreviated names of 
these objects are 2M0052$-$27, 2M0320$+$18, 2M1242$+$29 (M8), and 2M1504$-$23, 
2M1524$+$29, 2M1527$+$41, 2M1550$+$30 (M7.5). Their mean absolute magnitudes 
\filter{M$_J$}, \filter{M$_H$}, and \filter{M$_{K_s}$} are 
\magnit{11}{45}, \magnit{10}{76}, \magnit{10}{28} for the M8 sources, and 
\magnit{11}{11}, \magnit{10}{48}, \magnit{10}{07} for the M7.5 sources. 

Using these reference data, we can estimate the distances to the new sources. 
For LP\,775-\,31, we obtain 
$d_J$=6.2\,pc, $d_H$=6.4\,pc, and $d_{K_s}$=6.5\,pc, 
and for LP\,655-\,48, we obtain 
$d_J$=8.2\,pc, $d_H$=8.0\,pc, and $d_{K_s}$=7.9\,pc.
Making a conservative assumption of $\pm$\magnit{0}{4} accuracy in the
absolute magnitudes, we therefore estimate spectroscopic distances of 
$6.4\pm1.4$\,pc for LP\,775-\,31 and $8.0\pm1.6$\,pc for LP\,655-\,48.
It is worth noting that there is a significant possibility ($\sim$20\%)
of either object being a binary system (Close \etal{} \cite{close02}),
in which case the distance given here would be an underestimate: in the 
worst case of an equal mass binary, the true distance would be $\sim$1.4
times larger.

\section{Conclusions}
By combining data from the NLTT, 2MASS, and SSS, we have discovered two very 
nearby late M~dwarfs: LP\,775-\,31 has a spectral type of M8 and lies at a 
distance of $6.4\pm1.4$\,pc, while LP\,655-\,48 has a spectral type of M7.5 
and lies at $8.0\pm1.6$\,pc. Our spectroscopic distance estimates are 
roughly half those recently published by Cruz \& Reid (2002), namely
$11.3\pm 1.3$\,pc and $15.3\pm 2.6$\,pc, and need to be confirmed by checking 
the objects for a possible binary nature and by determining trigonometric 
parallaxes. If confirmed, they would be among the nearest stars to the Sun
and new benchmark M8 and M7.5 dwarfs, allowing detailed follow-up 
investigations. 

\begin{acknowledgements}
We have made use of data products from the SuperCOSMOS Sky Surveys at the 
Wide-Field Astronomy Unit of the Institute for Astronomy, University of 
Edinburgh, and from the Two Micron All Sky Survey, a joint project of the 
University of Massachusetts and IPAC, funded by NASA and the NSF\@.
The authors would like to thank ESO 3.6-m support team for useful
discussions, Nigel Hambly for advice on the use of SSS data, and
Darja Golikowa and Kai Schmitz for their help with the re-identification of 
NLTT stars in DSS and SSS images and with the cross-correlation of USNO A2.0, 
SSS, and 2MASS data. NL and MJM thank the EC RTN ``The 
Formation and Evolution of Young Stellar Clusters'' (HPRN-CT-2000-00155).
\end{acknowledgements}

\end{document}